\pdfoutput=1

\documentclass[letterpaper,english,reprint, aps]{revtex4-1}
\usepackage[T1]{fontenc}
\usepackage[latin9]{inputenc}
\usepackage{verbatim}
\usepackage{amsmath}
\usepackage{amssymb}
\usepackage{graphicx}
\PassOptionsToPackage{version=3}{mhchem}
\usepackage{mhchem}

\makeatletter

\usepackage{physics}

\makeatother

\usepackage{babel}
\begin{document}
\preprint{APS/123-QED}
\title{Quarter-filled Kane-Mele Hubbard model:\\
Dirac half-metals}
\author{S. Mellaerts\textsuperscript{1}, R. Meng\textsuperscript{1}, V.
Afanasiev\textsuperscript{1}, J.W. Seo\textsuperscript{2}, M. Houssa\textsuperscript{1,3}
and J.-P. Locquet\textsuperscript{1}}
\affiliation{\textsuperscript{1}Department of Physics and Astronomy, KU Leuven,
Celestijnenlaan 200D, 3001 Leuven, Belgium, \textsuperscript{2}Department
of Materials Engineering, KU Leuven, Kasteelpark Arenberg 44, 3001
Leuven, Belgium, \textsuperscript{3}Imec, Kapeldreef 75, 3001 Leuven, Belgium}
\email{simon.mellaerts@kuleuven.be}

\begin{abstract}
Recent experimental success in the realization of two-dimensional
(2D) magnetism has stimulated the search for new magnetic 2D materials
with strong magnetic anisotropy and high Curie temperature. One promising
subgroup of 2D magnetic systems are Dirac half-metals (DHM) which
have gained a lot of interest recently, as they host a high-temperature
quantum anomalous Hall effect (QAHE). This article discusses predictions
for intrinsic DHMs and identifies them as realizations of the Kane-Mele
Hubbard model at quarter filling. This proposed unification contributes
to a firmer understanding of these materials and suggests pathways
for the discovery of new DHM systems.
\end{abstract}
\maketitle

\section{\label{sec:Intro}Introduction}

The realization of magnetic semiconductors at room-temperature for
spintronic applications remains one of the biggest challenges in material
physics. Recently there have been many first-principle studies searching
to identify a magnetic semiconducting phase in two-dimensional (2D)
systems, including the prediction of various new concepts and materials,
such as spin-gapless semiconductors (SGS). The first SGS was proposed
by Wang \citep{WangSGS} in 2008 and can be considered as a material
being gapless in one spin channel and semiconducting/insulating in
the other. A distinction must be made between a parabolic and a linear
dispersion in the gapless spin channel, where the latter case leads
to a Dirac cone in one spin channel and materials which are often
called Dirac spin-gapless semiconductor (DSGS).

Importantly, as pointed out by Wang \emph{et al.} \citep{reviewDSGS},
there exists a difference between a DHM and DSGS. In both cases there
is a Dirac cone in one spin channel and a gap in the other, but in
the former case the Dirac cone is not necessarily located at the Fermi
level. However for convenience in this paper, we shall refer to these
together as the general group of DHMs. Furthermore, the group of DHMs
is further divided into $p$- and $d$-state DHMs, which refers to
the orbitals involved in the Dirac state. This work will mainly focus
on the intrinsic materials with $d$-state Dirac half-metallicity.

Since the first theoretical prediction of a DHM in a triangular ferrimagnet
\citep{FirstDHM} there have been many other suggestions of DHMs based
on first-principle calculations. For instance, the transition-metal
(TM) trihalides monolayers have been considered as an excellent system
for the realization of these DHM phases that all host a QAHE \citep{MBr3,NiCl3,MnX3,PtPdX3,FeX3}.
Moreover, V-group TM sesquichalcogenide with a honeycomb-kagome (HK)
lattice have also been predicted to achieve a DHM phase with room-temperature
QAHE \citep{V2O3,Nb2O3,Ta2S3}. However, a firm theoretical framework
of these DHMs and the room-temperature QAHE in these systems is still
lacking.

\ 

In this paper, we provide a theoretical framework based on the Kane-Mele
Hubbard (KMH) model that unifies the various predictions of DHMs,
with some of them hosting a room-temperature magnetic Chern insulator
by the existence of an atomic on-site spin-orbit coupling (SOC). Furthermore,
we discuss the universal features of these DHMs, where we elucidate
the importance of their multi-orbital structure.

\section{\label{sec:KMH}Kane-Mele Hubbard model}

As the quantum spin Hall effect (QSHE) in graphene and other 2D Xenes
(e.g. silicene) can be described by the Kane-Mele model \citep{KaneMele},
we argue that the realization of DHMs hosting a QAHE can be captured
by the inclusion of the electron correlations into the Kane-Mele model
- the KMH model - described by the hamiltonian

\begin{equation}
\begin{aligned}H & =-t\sum_{\langle ij\rangle,\alpha}\left(c_{i\alpha}^{\dagger}c_{j\alpha}+h.c.\right)\\
 & -i\lambda_{SO}\sum_{\langle\langle ij\rangle\rangle,\sigma}\left(c_{i\alpha}^{\dagger}\nu_{ij}\sigma_{\alpha\beta}^{z}c_{j\beta}+h.c.\right)\\
 & +U\sum_{i}n_{i\uparrow}n_{i\downarrow}
\end{aligned}
\end{equation}

where $t$ is the hopping parameter, $\lambda_{SO}$ the spin-orbit
coupling parameter, $U$ the Hubbard on-site parameter, $c_{i\alpha}^{\dagger}$
($c_{i\alpha}$) the electron creation (annihilation) operator at
site $i$ with spin state $\alpha\in\{\uparrow,\downarrow\}$ on the
honeycomb lattice, $n_{i\alpha}$ the number density operator, $\sigma^{z}$
a Pauli matrix, $\langle ij\rangle$ and $\langle\langle ij\rangle\rangle$
denotes resp. nearest and next-nearest neighbour hopping, and $\nu_{ij}=\pm1$
depending on whether the hopping path bends to the right or left.

\ 

\begin{figure}[t]
\includegraphics[scale=0.45]{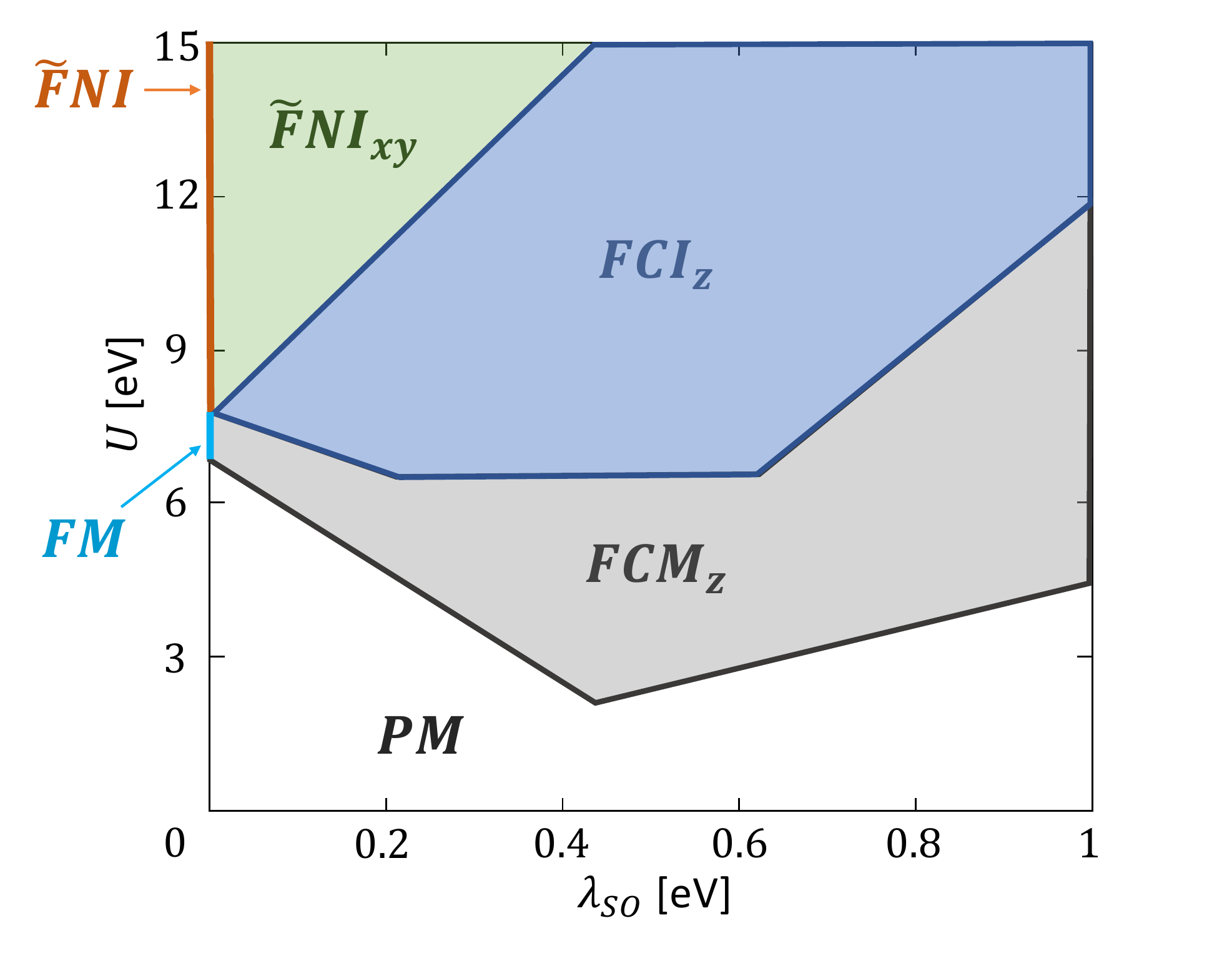}

\caption{The simplified phase diagram of the Kane-Mele-Hubbard model at quarter
filling on the honeycomb lattice where the hopping parameter is taken
to be $t=1$ eV, with the on-site Coulomb repulsion on the $y$-axis
and on the $x$-axis the intrinsic SOC. The phases present in the
phase diagram are: paramagnetic metal ($PM$), FM metal ($FM$), FM
normal insulator ($FNI$), FM Chern metal ($FCM$), FM Chern insulator
($FCI$), and other related phases with $F$ or $\tilde{F}$ to indicate
FM order with or without inversion symmetry and subscripts $xy$ or
$z$ representing the direction of the magnetic order. Adapted from
\citep{KHM}.}

\label{fig:Phasediagram}
\end{figure}

This model has been studied intensively in the half-filled situation
\citep{HalfKHM1,HalfKHM2,HalfKHM3}, however only a few studies explored
the case of quarter filling. Recently, this model was solved within
the mean-field approximation and its phase diagram was obtained \citep{KHM},
shown simplified in Fig. \ref{fig:Phasediagram}. In the case of non-zero
$\lambda_{SO}$ and small $U$, the exchange splitting will not be
sufficiently strong to establish a spin-polarized Dirac cone, and
the eventual gap opening by SOC at the Dirac point will be closed
by the overlapping of the spin-minority bands. Under those conditions
a paramagnetic metal ($PM$) or ferromagnetic Chern metal phase ($FCM_{z}$)
is stabilized. Upon further increasing $U$ a magnetic Chern insulating
phase ($FCI_{z}$) with strong magnetic anisotropy is established.
This situation can be understood as the cooperative effect of strong
$U$ and $\lambda_{SO}$, where the former results in a large exchange
splitting, whereas the latter provides the strong magnetic anisotropy
\citep{SOC_anisotropy} and opens a gap at the Dirac point. On further
increasing $U$, a trivial Mott insulating phase ($\widetilde{F}NI_{xy}$)
with broken inversion symmetry will appear, however in contrast to
a typical antiferromagnetic Mott insulator, an in-plane FM order is
established as explained in \citep{inplaneAFM}.

\section{\label{sec:Unification}Unification of DHM}

\begin{figure*}
\includegraphics[scale=0.68]{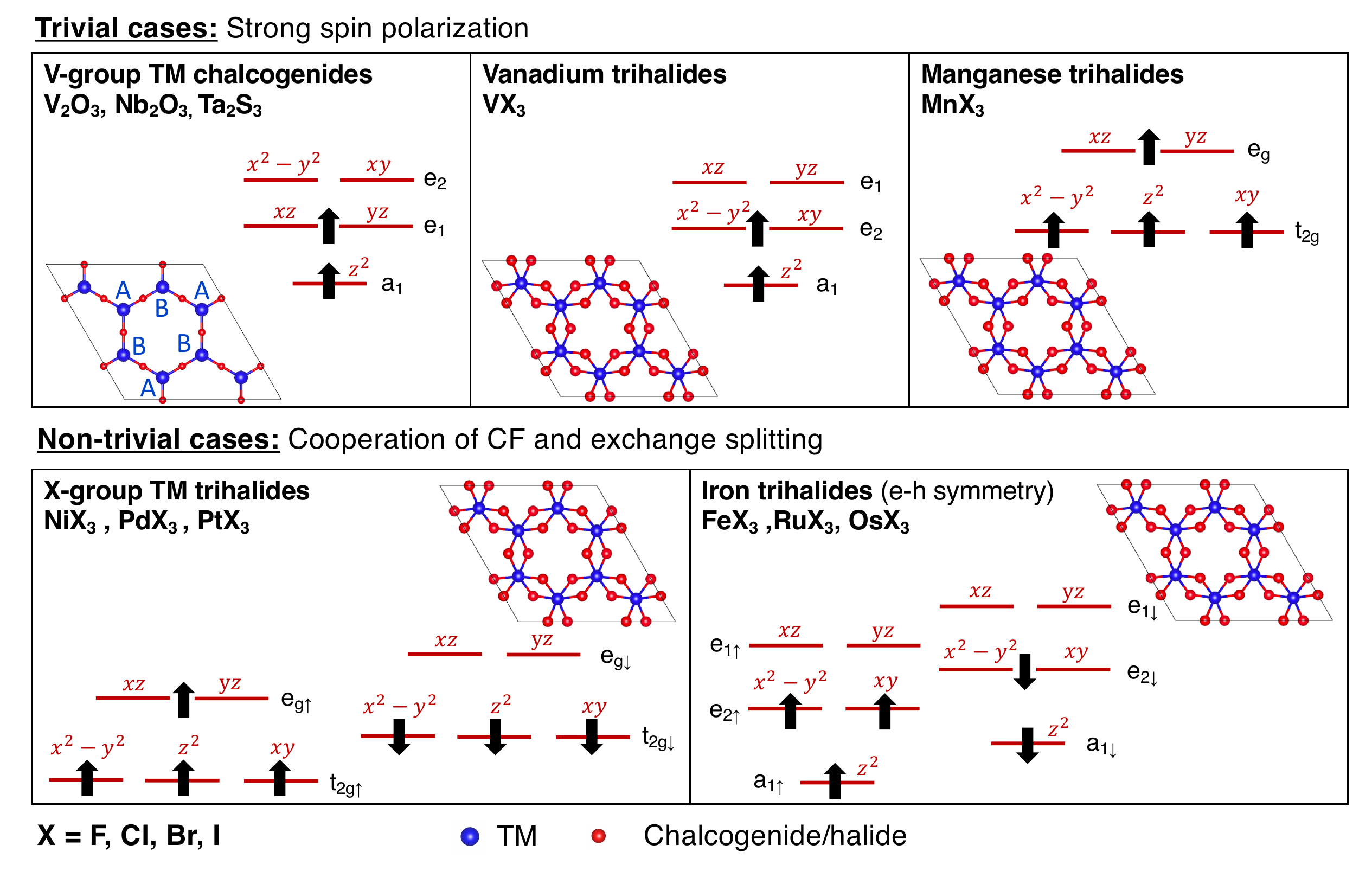}

\caption{Unification scheme of the Dirac half-metals based on the quarter-filled
KMH model. The lattice structure (with sublattices $\tau=A,B$) for
each realization of quarter-filled $d$-orbital level is depicted
with the blue atoms representing the TM and the red atoms the chalcogenides/halides.
The unification scheme is further divided by trivial cases where a
strong $U$ results in a strong exchange splitting, whereas in the
non-trivial cases there is the cooperation of a CF and exchange splitting
to satisfy the quarter-filled level condition.}

\label{fig:Unification}
\end{figure*}

To our knowledge, the DHM phases hosting a QAHE can be divided into
two classes: V-group TM sesquichalcogenide monolayers with HK lattice
structure and the honeycomb TM trihalide monolayers, with their respective
lattice structure . shown in Fig. \ref{fig:Unification}(top). The
former group can be further subdivided by its crystal field (CF) environment
as either octahedral or distorted octahedral. This gives rise to three
types of CF splitting situations, as illustrated in Fig. \ref{fig:Unification}(top).
In all these situations the TM forms a honeycomb structure, which
is crucial for the existence of a Dirac cone \citep{existence}. 

The most simple realization of quarter filling of the $d$ orbital
levels is the situation of strong $U$, where exchange splitting between
the spin-majority and minority channels is sufficiently large such
that only one spin channel remains close to the Fermi level. For the
distorted octahedral case, two remaining electrons on the TM ion after
bonding with the halide gives rise to quarter filling of the $e_{2}$
orbital level, which is realized in $\ce{\ce{{VBr_{3}}}}$ \citep{MBr3}
due to the two-fold degeneracy of this \emph{$(d_{xy},d_{x^{2}-y^{2}})$}
level. In the case of octahedral CF splitting, there would be quarter
filling of $e_{g}$ for four remaining electrons on the TM ion, which
happens for $\ce{\ce{{MnX_{3}}}}$ with $\ce{\ce{{X}}}$ = $\ce{\ce{{F}}}$,
$\ce{\ce{{Cl}}}$, $\ce{\ce{{Br}}}$, $\ce{\ce{{I}}}$ \citep{MnX3}.
Finally, in the HK $\ce{\ce{{M_{2}X_{3}}}}$ with trigonal CF there
will be a quarter filled $e_{1}$ orbital when there are two remaining
electrons on the TM ion after bonding with the chalcogenides. This
situation has been established in $\ce{\ce{{V_{2}O_{3}}}}$ \citep{V2O3},
$\ce{\ce{{Ta_{2}S_{3}}}}$ \citep{Ta2S3} and $\ce{\ce{{Nb_{2}O_{3}}}}$
\citep{Nb2O3}. All of the above mentioned systems have been predicted
to be a DHM phase hosting QAHE. These situations can be understood
as the realization of the Haldane model, where the strong $U$ and
SOC result in the formation of a strong magnetic moment which destroys
one spin channel of the Kane-Mele model, leaving one QAH state.

\ 

These are the most trivial situations that confirm part of the phase
diagram in Fig. \ref{fig:Phasediagram}. However, this KMH tight-binding
(TB) model does not take into account the multi-orbital level nature
of these compounds, which allows the realization of many more situations
where the combination of CF splitting and exchange splitting allows
the quarter filling of the orbital levels at much lower $U$. In this
way, TM with higher electron occupation of the $d$-orbitals can also
stabilize a DHM phase. For example, $\ce{\ce{{NiCl_{3}}}}$ \citep{NiCl3},
$\ce{\ce{{PtBr_{3}}}}$ and $\ce{\ce{{PdBr_{3}}}}$ \citep{PtPdX3}
where the $4d$ and $5d$ have a slightly lower $U$ but where the
strong CF splitting allows the situation of six remaining electrons
to fill the $t_{2g}$ level and one remaining electron that leads
to quarter filling of the $e_{g}$ level, as illustrated at the bottom
of Fig. \ref{fig:Unification}. Moreover, due to the presence of particle-hole
symmetry, the realization of a three-quarter filled $d$-orbital is
equivalent. This situation was predicted in iron trihalides $\ce{\ce{{FeX_{3}}}}$
\citep{FeX3}. Again these type of systems have also been predicted
to be DHM phases with a SOC-induced QAHE.

Furthermore, some of these systems (e.g. $\ce{\ce{{VBr_{3}}}}$ and
$\ce{\ce{{FeX_{3}}}}$) have also been predicted to become a Mott
insulator by including a sufficiently large $U$ ($>0.5$ eV). We
relate these systems to the KMH model by arguing that the quarter-filled
orbital level consists of the in-plane $(d_{xy},d_{x^{2}-y^{2}})$
orbitals, which will have a larger overlap leading to a larger hopping
term $t$ . Consequently, rescaling the phase diagram by $t$ should
make the Mott insulating phase ($\widetilde{F}NI_{xy}$) accessible
at lower $U$.

The various realizations of the quarter-filled KMH model are summarized
in Fig. \ref{fig:Unification}, which are all predicted to be DHMs
hosting a QAHE. They can be classified according to a scheme where
quarter-filling of the $d$-orbital level is established. Identifying
this crucial ingredient of quarter-filling provides the opportunity
to search for other DHMs. For example, we expect that materials with
TM and halide/chalcogenide in the same chemical group can also yield
the same quarter-filled orbitals. Nevertheless, this does not mean
that this quarter-filled KMH model is inclusive in the sense that
all intrinsic $d$-state DHMs should be the realization of this model.
Moreover, this simple TB model has also certain shortcomings. The
most important one is the prediction of the magnetic ground state
and magnetic anisotropy where CF splitting, orbital orientations and
exchange interactions are critical and the TB model might not capture
these accurately.%
{} Nonetheless, as will be shown in Sec. \ref{sec:Universal} this model
gives insights on how the cooperative effect of SOC and electron correlations
can stabilize high-temperature magnetic Chern insulators.

\begin{figure*}[t]
\includegraphics[scale=0.72]{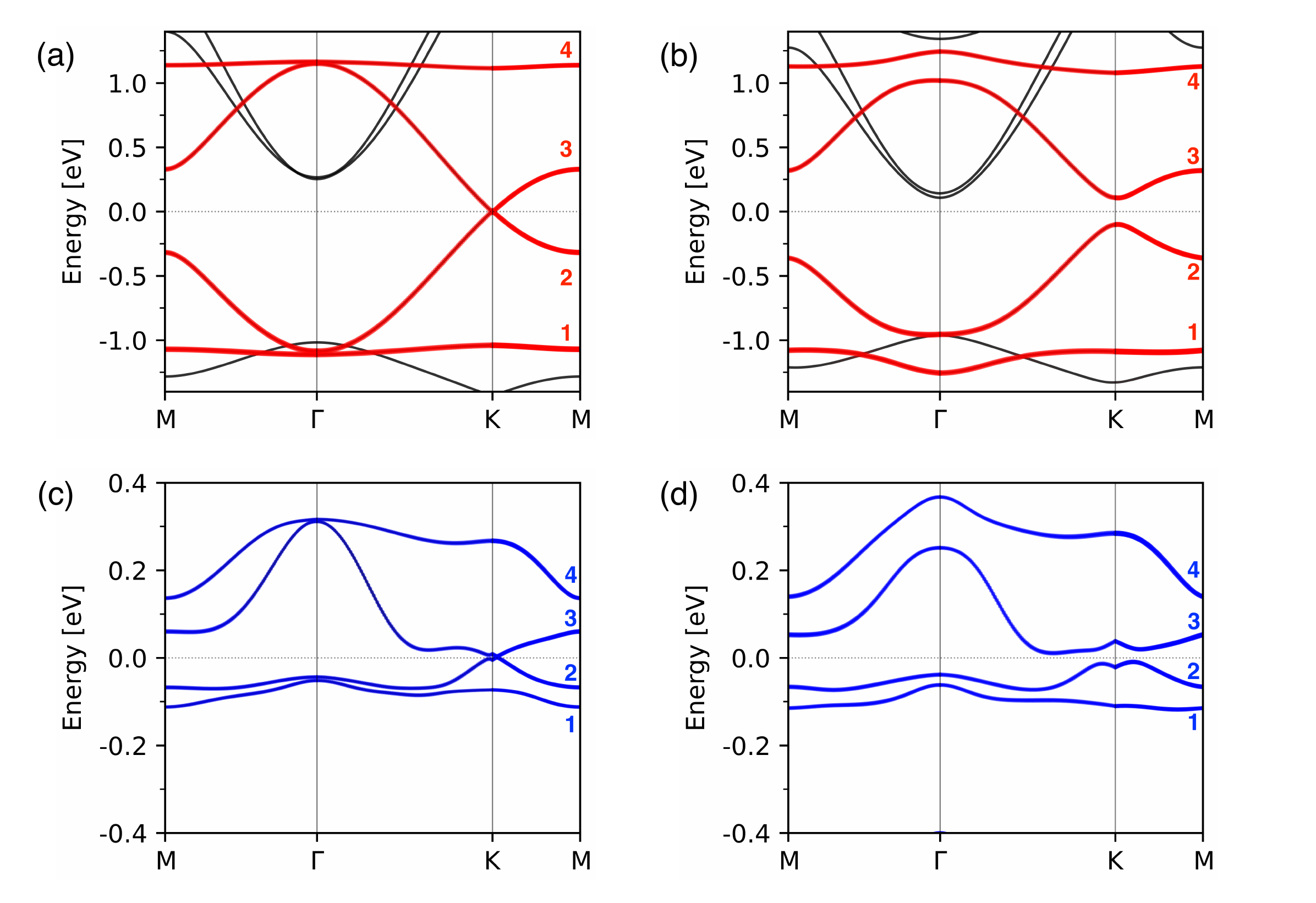}\caption{The DFT calculated band structure of DHMs with the four low-energy
bands indicated by red/blue. The band structure of (a-b) $\ce{V_{2}O_{3} }$
and (c-d) $\ce{MnBr_{3}}$ without and with SOC respectively.}

\label{fig:bands}
\end{figure*}

\section{\label{sec:Universal}Universal features of DHMs}

In this section, a more detailed study of the low energy physics and
the realized QAHE is given. We identify the common features and unify
them within a TB model adopted in earlier studies on the $(p_{x},p_{y})$
honeycomb lattice systems \citep{atomicOnsiteSOC,multiorbitalHoneycomb}.
Before discussing the universal features, it is also important to
make a further distinction between two systems within the unification
scheme in Fig. \ref{fig:Unification}, depending on the type of quarter-filled
degenerate orbital levels involved. In the case of quarter-filled
$(d_{xz},d_{yz})$-orbitals, there will be a $\pi$-bond between the
TMs with only a small orbital overlap and thus small hybridization
with the $p$-orbitals of the chalcogenides/halides. In this case,
it can be expected that a TB model will be fairly accurate capturing
most of the features. However, in the other case of quarter-filled
$(d_{xy},d_{x^{2}-y^{2}})$-orbitals, a strong in-plane hybridization
with the ligand orbitals is expected, making a TB approach a less
appropriate description.

\subsection{Topological band structure}

The two-fold degenerate $d$-orbitals that are quarter-filled will
be responsible for the low-energy physics of the system. The two-fold
degeneracy of the orbitals together with the sublattice symmetry gives
rise to a total of four bands close to the Fermi level $E_{F}$, as
shown in Fig. \ref{fig:bands}. Two bands are dispersive and form
a Dirac cone at the $K$ ($K'$) point (band $2$ and $3$), while
the other two bands are (nearly-)flat (band $1$ and $4$) forming
a degeneracy with the dispersive bands at the $\Gamma$ point. As
examples of the low-energy band structures of TM sesquichalcogenide
and TM trihalides, the band structure of resp. $\ce{V_{2}O_{3} }$
and $\ce{MnBr_{3}}$ are shown in Fig. \ref{fig:bands}. These band
structures were calculated by density functional theory (DFT) (See
Supplemental Material). 

\ 

This four-band low-energy physics of the DHMs is very similar to the
low-energy physics of the $(p_{x},p_{y})$-orbital honeycomb lattice
which has been studied intensively in the context of ultracold-atom
optical lattices \citep{FB1,FB2,pxpy_fwaveparing}. Therefore, these
TB calculations and their results can be used for the description
of these DHM phases. Here, only the main results of this four-band
model are formulated, the full calculation can be found in \citep{multiorbitalHoneycomb}.
Neglecting the spin degree of freedom, a four-component basis can
be constructed,

\begin{equation}
\psi_{\tau\sigma}=\left(d_{A,+},d_{B,+},d_{A,-},d_{B,-}\right)
\end{equation}

with $d_{\tau,\pm}^{\dagger}=\frac{1}{\sqrt{{2}}}(d_{\tau,xz}^{\dagger}\pm id_{\tau,yz}^{\dagger}$)
or $d_{\tau,\pm}^{\dagger}=\frac{1}{\sqrt{{2}}}(d_{\tau,xy}^{\dagger}\pm id_{\tau,x^{2}-y^{2}}^{\dagger})$
the orbital angular momentum $L_{z}$ eigenstates with resp. eigenvalues
$L_{z}=\pm1$ and $L_{z}=\pm2$, and with sublattice component $\tau=A,B$.
In other words, the doublet of orbital angular momentum and the sublattice
structure give rise to two pseudospin degrees of freedom, which are
only conserved at certain high-symmetry points. For simplicity, we
will restrict the discussion to the case of a four-band model for
the orbital angular momentum eigenstates $d_{\tau,\pm}=\frac{1}{\sqrt{{2}}}(d_{\tau,xz}^{\dagger}\pm id_{\tau,yz}^{\dagger})$
with $L_{z}=\pm1$. However, this discussion can be easily generalized
to the other situation with $L_{z}=\pm2$. 

\ 

At the $K$ point, the eigenstates of the two middle bands are orbital
eigenstates with $L_{z}=\pm1$ where the spin-conserving on-site SOC
will lift this orbital degeneracy, opening an energy gap equal to
the eigenenergy difference between the SOC term $s_{z}L_{z}$ with
$L_{z}=\pm1$. As a result, the eigenstate of the band $2$ will be
the orbital eigenstate $d_{xz}+id_{yz}$ with $L_{z}=+1$ and eigenenergy
$-\lambda_{SO}$ whose wave function is completely located on the
$B$ sublattice, while band $3$ has orbital eigenstate $d_{xz}-id_{yz}$
with $L_{z}=-1$ and eigenenergy $+\lambda_{SO}$ completely located
on sublattice $A$. Hence, the on-site SOC opens a topologically non-trivial
energy gap of $2\lambda_{SO}$ in the presence of sublattice symmetry.
This analysis can be easily generalized to the $K'$ point.

\ 

On the other hand, at the $\Gamma$ point, the hamiltonian $H(k)$
preserves all rotation symmetries i.e. the SOC term $s_{z}L_{z}$
commutes with $H(k)$. Consequently, the bands near the $\Gamma$
point will have eigenstates which are the superposition of wave functions
located on both $A$ and $B$ sublattices. Nonetheless, band $1$
has orbital eigenstate with $L_{z}=-1$, while the eigenstate of band
$2$ is an $L_{z}=+1$ eigenstate. Therefore, similar to the $K$
($K'$) points, the on-site SOC opens a non-trivial energy gap equal
to the eigenenergy difference of the SOC term, $2\lambda_{SO}$. The
topological nature of this gap opening at $\Gamma$ away from $E_{F}$
has been largely unrecognized in first-principles calculations.

\subsection{Large energy gap and cooperation of $U$ and SOC }

The realization of the QAHE in DHMs arises from the cooperative effect
of strong electron correlations and SOC, as stated in many first-principles
studies on DHMs. However, the true origin of this relatively large
energy gaps remains not well understood. Two excellent theoretical
studies on $(p_{x},p_{y})$ honeycomb lattice systems have clarified
the observation of the large energy gaps opened by an intrinsic \emph{atomic}
SOC \citep{atomicOnsiteSOC,multiorbitalHoneycomb}. In the literature,
there have been two common mechanisms giving rise to topological band
gaps: first the band inversion mechanism of two bands with different
orbital character and second the next-nearest-neighbor (NNN) intrinsic
SOC as in the Kane-Mele model. This latter mechanism, often at play
in 2D topological materials, is typically a (second-order) NNN hopping
process which only generates a relatively small energy gap. Therefore
one seeks other topological materials where a sizeable SOC-induced
gap can be found.

\ 

In graphene, the leading-order effective SOC is a second-order NNN
process illustrated by the following sequence:

\begin{equation}
\begin{aligned}\ket{p_{z\uparrow}^{A}}\xrightarrow{SOC}\ket{p_{+\downarrow}^{A}}\xrightarrow{V}\ket{s_{\downarrow}^{B}}\xrightarrow{V}\ket{p_{+\downarrow}^{A}}\xrightarrow{SOC}\ket{p_{z\uparrow}^{A}}\\
\ket{p_{z\downarrow}^{B}}\xrightarrow{SOC}\ket{p_{-\downarrow}^{B}}\xrightarrow{V}\ket{s_{\uparrow}^{B}}\xrightarrow{V}\ket{p_{-\uparrow}^{B}}\xrightarrow{SOC}\ket{p_{z\downarrow}^{B}}
\end{aligned}
\end{equation}

with orbital angular momentum eigenstate $\ket{p_{\pm}^{\tau}}=p_{x}^{\tau}\pm ip_{y}^{\tau}$
, $SOC$ indicating the \emph{intrinsic} atomic on-site SOC process,
and $V$ the nearest-neighbor direct hopping amplitude. This whole
NNN hopping process involves an on-site SOC with two spin-flip $\ket{p_{z\uparrow}^{A}}\xrightarrow{SOC}\ket{p_{\pm\downarrow}^{A}}$
processes from the $\pi$-band to the $\sigma$-band, resulting in
no net spin-flip effective SOC. As the intrinsic atomic SOC appears
twice, the effective SOC will be a second-order process, making it
a relatively weak effect. The importance of the multi-orbital nature
became clear in the slightly larger SOC-induced gap opening in silicene
and other Xenes \citep{houssa2015silicene,houssaGermanene,houssa2016stanene,Xenes},
which have a low-buckled structure. In this case, the leading-order
effective SOC is a first-order NNN process:

\begin{equation}
\begin{aligned}\ket{p_{z\sigma}^{A}}\xrightarrow{V}\ket{p_{-\sigma}^{B}}\xrightarrow{\mp SOC}\ket{p_{-\sigma}^{B}}\xrightarrow{V}\ket{p_{z\sigma}^{A}}\\
\ket{p_{z\sigma}^{B}}\xrightarrow{V}\ket{p_{-\sigma}^{A}}\xrightarrow{\pm SOC}\ket{p_{-\sigma}^{A}}\xrightarrow{V}\ket{p_{z\sigma}^{B}}
\end{aligned}
\end{equation}

where the low-buckled structure allows the coupling between the out-of-plane
oriented $p_{z}^{A}$ orbital and the in-plane oriented $p_{\pm}^{B}$
orbitals. Hence, the electrons in $p_{z}^{A}$ will hop to the nearest-neighbor
$p_{\pm}^{B}$ where they will experience the intrinsic atomic SOC,
followed by a hopping to another nearest-neighbor $p_{z}^{A}$. Consequently,
the effective SOC is a first-order NNN process, making it a slightly
stronger effect when the buckling is sufficiently large to have a
strong coupling (large $V$) between $p_{z}^{A}$ and $p_{\pm}^{B}$
. 

\ 

The degenerate nature of the $d$-orbital levels in the DHM phases
allows for the existence of an on-site atomic SOC without any nearest-neighbor
hopping. In other words, the leading-order effective SOC processes
are

\begin{equation}
\begin{aligned}\ket{d_{\pm\uparrow}^{\tau}}\xrightarrow{\pm\lambda_{SO}}\ket{d_{\pm\uparrow}^{\tau}}\\
\ket{d_{\pm\downarrow}^{\tau}}\xrightarrow{\mp\lambda_{SO}}\ket{d_{\pm\downarrow}^{\tau}}
\end{aligned}
\end{equation}

which are a first-order on-site processes. Intuitively we can think
of a multi-orbital structure at each site allowing for intra-orbital
hopping where the electrons will experience a strong atomic on-site
SOC. Consequently, this atomic on-site SOC can result in large energy
gap magnitudes (e.g. $X$-hydride/halide monolayer systems have energy
gaps up to $0.65$ eV \citep{atomicOnsiteSOC}). Moreover, within
the understanding of an energy gap opened by an atomic SOC, the cooperative
behavior of the SOC and electron correlations ($U$) becomes evident;
the Hubbard correction $U$ will result in the localization of the
electrons yielding a stronger atomic SOC effect. In terms of perturbation
theory, $U$ will force the orbital wave functions into more atomic-like
wavefunctions over which the expectation value is taken when evaluating
the SOC perturbation.

\section{Conclusions}

The obtained results provide a deeper understanding of a new type
of DHM phases. We first unified the predictions of the intrinsic $d$-state
DHMs by a simple TB model, the Kane-Mele Hubbard model at quarter
filling. This identification provides a guide for the prediction of
new DHMs, but also illustrates that DHMs might host many other phases
under external conditions. In addition, some physical descriptions
were given to obtain a better understanding on the cooperative effect
of electron correlations and SOC. We elucidated the importance of
the multiorbital structure for the existence of a strong\emph{ }on-site
SOC and large energy gap, which is required for the room-temperature
application of DHMs hosting a QAHE.

\section*{Acknowledgements}

Part of this work was financially supported by the KU Leuven Research
Funds, Project No. KAC24/18/056 and No. C14/17/080 as well as the
Research Funds of the INTERREG-E-TEST Project (EMR113) and INTERREG-VL-NL-ETPATHFINDER
Project (0559). Part of the computational resources and services used
in this work were provided by the VSC (Flemish Supercomputer Center)
funded by the Research Foundation Flanders (FWO) and the Flemish government.

\section*{Supplementary Material}

All spin-polarized calculations were carried out within DFT, as implemented
in Vienna \emph{ab initio }simulation package (VASP) \citep{VASP}.
The generalized gradient approximation in the form of Perdew-Burke-Ernzerhof
(PBE) \citep{PBE} was used as exchange-correlation functional. The
interactions between the electrons and ions were described by the
projected-augmented-wave (PAW) pseudopotentials \citep{PAWmethod},
and an energy cutoff of $550$ eV was adopted. For the structural
relaxation, a force convergence criterion of $0.005$ eV/$\AA$
was used with the Brillouin zone (BZ) sampled by a $12\times12\times1$
$\Gamma$-centered $k$-point mesh, with a vacuum space of $17\,\AA$
adopted along the normal of the lattice plane. To account for the
localized nature of the $3d$ electrons of the TM cation a Hubbard
correction $U$ is employed within the rotationally invariant approach
proposed by Dudarev et al. \citep{Dudarev}, where $U_{\text{eff}}=U-J$
is the only meaningful parameter. A Hubbard correction of $3$ eV
and $4$ eV is adopted for $\ce{V_{2}O_{3} }$ and $\ce{MnBr_{3}}$
respectively. The electronic self-consistent calculations were performed
with total energy convergence criterion of $10^{-6}$ eV with the
BZ sampled by a denser $24\times24\times1$ $\Gamma$-centered $k$-point
mesh. The atomic structures were visualized by the VESTA program \citep{VESTA}.

\bibliographystyle{unsrt}
\bibliography{referencesDHM}

\end{document}